\documentstyle[aps,prb,twocolumn]{revtex}
\input epsf
\draft
\begin{document}
\preprint{NIRIM-MA-97-01}
\title{ Electronic Structures of Sr$_{14-x}$Ca$_{x}$Cu$_{24}$O$_{41}$  }

\author{Masao Arai}

\address{
  National Institute for Research in Inorganic Materials, \\
  Tsukuba, Ibaraki 305, Japan \\
  }
\author{ Hirokazu Tsunetsugu}

\address{
  Institute of Applied Physics, University of Tsukuba, \\
  Tsukuba, Ibaraki 305, Japan
  }

\date{\today}
\maketitle
\begin{abstract}
  The electronic structures of Sr$_{14-x}$Ca$_{x}$Cu$_{24}$O$_{41}$
  are calculated within the local density approximation.  Around the
  Fermi energy there exist quasi-one-dimensional bands originated from
  the ladder and chain layers.  The nearest-neighbor inter-ladder
  hoppings are estimated to be 5--20\% of the intra-ladder ones.
  Possible effects of Ca substitution on electronic structures and charge
  distribution are also discussed.
\end{abstract}

The doped spin-ladder compound, Sr$_{14-x}$Ca$_{x}$Cu$_{24}$O$_{41}$, has
recently attracted wide attention as a possible new class of superconductor.
Superconductivity was observed at $x$=13.6 under high pressure 3.5$\sim$4
GPa.\cite{akimitsu}
The insulating phase has a finite spin gap,
and the superconductivity is expected to be driven by the spin-liquid
ground state, as theoretically predicted in Ref.\ \onlinecite{super}.

The superconductivity appears upon substituting isovalent Ca for Sr,
in addition to applying high pressure, whereas the formal Cu valence
stays constant, 2.25 for all $x$.  Considering this compound contains
chain layers as well as ladder layers, this indicates that the local
charge distribution may change with Ca substitution and with pressure,
and that superconductivity needs a special charge distribution.
Various experiments\cite{uchida,kitaoka} and model
calculations\cite{mkato,mizuno} have shown that most of the holes
exist on the chains for Sr$_{14}$Cu$_{24}$O$_{41}$ and that the Ca
substitution transfers the holes to ladder layers. In the present
paper, we study the electronic structures and charge distribution of
Sr$_{14-x}$Ca$_{x}$Cu$_{24}$O$_{41}$ within the local density
approximation (LDA).\cite{LDA}

The crystal structures of Sr$_{14}$Cu$_{24}$O$_{41}$ and related
compounds were reported in
Refs.~\onlinecite{kato1,LaCa-struc,kato2,Sr-struc,Sr-ukei}.  They are
composed of two subsystems, Sr$_2$Cu$_2$O$_3$ (ladders and Sr) and
CuO$_2$ (chains).  The chemical formula determined by structure
analysis\cite{Sr-ukei} is (Sr$_2$Cu$_2$O$_3$)(CuO$_2$)$_y$ with $y =
1.436$.  Due to interactions between the two subsystems, atomic
positions of each subsystem are modulated by the periodicity of the
other.  Their structures were analyzed by assuming a large unit cell
with $y$ approximated by a rational number,\cite{kato1,LaCa-struc} or
by the super space-group technique.\cite{kato2,Sr-ukei} If $y$ is
chosen as ${7 \over 5}$ or ${10 \over 7}$, the chemical formula
becomes Sr$_{10}$Cu$_{17}$O$_{29}$ and Sr$_{14}$Cu$_{24}$O$_{41}$,
respectively.

We have calculated the electronic structures for
M$_{10}$Cu$_{17}$O$_{29}$ and M$_{14}$Cu$_{24}$O$_{41}$ (M = Sr or
Ca).  As a starting point, the structure modulation is ignored for
simplicity.  Possible effects of the modulation will be briefly
discussed later. The symmetry of M$_{10}$Cu$_{17}$O$_{29}$ is chosen
as face-centered orthorhombic F222 following Ref.~\onlinecite{kato1}.
The $a$-axis is perpendicular to the ladders and chains inside layers,
the $b$-axis perpendicular to stacking layers, and the $c$-axis
parallel to the chains and ladders.  The lattice parameters are listed
in Table~I. $c_1$ ($c_2$) is the fundamental period of the ladder
(chain) part along the $c$-axis.

Even with the rational approximation, the crystal structure of
M$_{10}$Cu$_{17}$O$_{29}$ is still complicated and the unit cell
contains 56 atoms.  Such a complicated structure is challenging for
the {\it ab initio} calculation.  We used the
linear-muffin-tin-orbital (LMTO) method\cite{LMTO} with the
atomic-sphere approximation, since this is suitable for a large unit
cell.  In addition to 56 atomic spheres, 44 empty spheres are
inserted around Sr and CuO$_2$ chains.  The positions and size of the
empty spheres are optimized by the method explained in
Ref.~\onlinecite{JA}. 
We have performed the self-consistent calculations with 63 {\it k\/}
points in the irreducible Brillouin zone for
M$_{10}$Cu$_{17}$O$_{29}$ and 21 {\it k\/} points for
M$_{14}$Cu$_{24}$O$_{41}$.

First, we show the total and partial density of states (DOS) in
Fig.~1.  The Fermi energy is set at $E_F$=0.  The calculated DOS of
Sr$_{10}$Cu$_{17}$O$_{29}$ and Sr$_{14}$Cu$_{24}$O$_{41}$ are similar
to each other.  This indicate that the small unit cell is enough to
investigate the qualitative features.  These compounds are calculated
as a metal, leading to finite DOS at $E_F$.  To reproduce experimentally
observed insulating behavior, it would be necessary to take into
account electron correlations along with possible charge ordering. The
states in the region $-$7$\sim$2eV are mainly composed of the Cu $d$-
and O $p$-orbitals.  Both Cu and O partial DOS distribute broadly in
this energy region, which indicates strong hybridization of these orbitals.
The Sr $s$-orbitals slightly hybridize with the Cu $d$- and O
$p$-orbitals, while their main weights appear at higher energy.

The width of the valence band is consistent with the
photoemission experiments.\cite{T.T}  The two broad peaks observed in
the photoemission spectrum at binding energy 3eV and 5.5eV are
actually seen in the present calculation. The former peak may be
assigned to the peaks of the total DOS at $-$2$\sim$$-$3eV, and the
latter peak around $-$5eV. However, the position of Fermi energy is about
1eV lower in the calculation.  This may be because the Fermi energy is
not correctly calculated for this metallic ground state.

The chain and ladder layers have different structures of partial DOS
around $E_F$.  The chain DOS forms a band with width of about 1eV,
separated from the main valence bands with an energy gap of 1eV.  On
the other hand, the ladder DOS is smaller than the chain DOS and
distributes in a wider energy region. The peaks at 0.1eV and 1.5eV in
the ladder DOS are due to the edge singularity of two
quasi-one-dimensional bands, as we shall explain later.  The
inter-layer hybridization between ladders and chains seem to be
small, since the peak positions of their partial DOS are not
correlated.

We show the energy bands near $E_F$ in Fig.~2.  The bands are most
dispersive along the symmetry line $\Gamma$--Z, which is parallel to
chains and ladders, and then along $\Gamma$--X (perpendicular to
chains and ladders in layers).  The small dispersion perpendicular to
layers ($\Gamma$--Y) indicates that the interlayer hoppings are weak.
We identify the character of each state by the weights of various
atomic orbitals.  Most of states near $E_F$ have their weights only on
either chain layers or ladder layers.  This verifies small interlayer
hoppings, consistent with the dispersionless character along
$\Gamma$--Y.  We also find that the chain bands near $E_F$ are mainly
composed of the antibonding combination of Cu $d_{xz}$-orbitals and O
$p_x$- and $p_z$-orbitals, while the ladder bands are composed of Cu
$d_{x^2-z^2}$-orbitals and O $p_x$- and $p_z$-orbitals.

We find that the calculated energy bands could be simply understood in
the following way.  To see that, we plot in Fig.~3 the chain and
ladder bands separately in the {\em extended} Brillouin zone.  The
states with more than 15\% of their weights on Cu
$d_{x^2-z^2}$-orbitals in the ladders are shown with large dots in
Fig.~3 (a) and (b).  The difference is that those in (a) and (b) have
an even and odd parity, respectively, concerning the mirror symmetry
perpendicular to the $a$-axis.  They both have the approximate
periodicity of ${20\pi \over c}={4\pi \over c_1}$, corresponding to
the fundamental ladder period.  Thus, it is reasonable to interpret
that these bands are generated by the folding of an energy band
between ${\bf k}$=(0,0,0) and (0,0,${2\pi \over c_1}$).  The band can
be fitted by the following tight-binding dispersion:
\begin{eqnarray}
  &\epsilon_{\bf k} &= \varepsilon_0 - 2 t_1 \cos( k_z c_1)
  - 2 t_2 \cos(2 k_z c_1) \nonumber  \\
  && - \left[ 
           4 t^{\perp}_1 \cos( {\textstyle \frac{1}{2}}k_z c_1) +
           4 t^{\perp}_2 \cos({\textstyle \frac{3}{2}}k_z c_1) 
       \right]
       \cos({\textstyle \frac{1}{2}}k_x a) . 
   \label{dispersion}
\end{eqnarray}
where $t_1$ and $t_2$ are the nearest and second nearest neighbor
hoppings between rungs in the ladders, while $t^{\perp}_1$ and
$t^{\perp}_2$ between adjacent ladders.  The results are summarized in
Table~II and the fitting is also shown in Fig.~3 (a) and (b).  With
the five parameters for each band, we can reproduce the overall
features.

The inter-ladder hoppings are 5--20\% of intra-ladder ones.  These
values suggest that the inter-ladder hoppings may not be negligible to
discuss the electronic structures and physical properties.
Experimentally, the anisotropy of resistivity $\rho_a/\rho_c$ is about
30 at temperature 100K for
Sr$_3$Ca$_{11}$Cu$_{24}$O$_{41}$,\cite{motoyama} larger than the
present estimation, and it also has a large temperature dependence.
Therefore it is natural to expect that the strong correlation effects
such as hole pairing enhance the anisotropy, but we do not discuss it
further in this paper.  At high temperatures, the electron correlation
may not be important and the present band anisotropy is consistent
with the experiments.\cite{motoyama}

The chain bands near $E_F$ also have a pseudo one-dimensional
character, while their dispersions are more complex as shown in
Fig.~2(c).  Here large dots represent the states with more than 15\%
of their weights on Cu $d_{xz}$-orbitals in the chain layers.  We
again fitted the dispersion by Eq.\ (\ref{dispersion}).  The nearest
neighbor hoppings, $t_1$, along the chains are smaller than those for
the ladder bands.  This is because of the near 90$^{\circ}$ angle of
Cu-O-Cu bonds.\cite{mizuno} The second nearest neighbor hoppings,
$t_2$ are largest and roughly twice the nearest neighbor ones. The
sign and relative ratio of these two, $t_1$ and $t_2$, are consistent
with the semi quantitative estimation considering the spatial direction
of atomic orbitals.

Next, we discuss the effects of Ca substitution which corresponds to
positive chemical pressure. The unit cell volume decreases and the
lattice constant $b$, i.e., the distance of subsequent layers,
decreases most rapidly.  We have calculated the electronic structure
of fully substituted Ca$_{14}$Cu$_{24}$O$_{41}$ with the lattice
constants extrapolated from the available experimental
data.\cite{akimitsu} The internal atomic positions are fixed to the
same values with Sr$_{14}$Cu$_{24}$O$_{41}$.  The total and partial
DOS have similar features to those of Sr$_{14}$Cu$_{24}$O$_{41}$.  The
main change is a slight increase of the hybridization between Cu
$d$-orbitals and O $p$-orbitals.  The enhancement of the hybridization 
is actually found as the wider total band width and the larger 
tight-binding hopping integrals shown in Table~II.  
The increase is caused by the decrease 
of the atomic distance.

Let us now examine the hole distribution on the ladder and chain
layers, which is important to investigate the origin of
superconductivity under high pressure.
We use the occupation ratio of the chain bands near $E_F$ as a
measure of the hole concentration.  The bands are composed
of the antibonding combination of Cu $d_{xz}$- and O $p_{x}$- and
$p_{z}$-orbitals.  Setting the O valence to $-$2, we determine that the
valence of chain Cu would be +1 if these bands are fully occupied, and +3
if completely empty.  This is exact if the interlayer hybridization
were absent and is expected to hold qualitatively in the present case,
too.

With this assumption, we have calculated the chain Cu valence as
$P_c=3-2p$, where $p$ is the occupation ratio of the chain DOS to the
whole number of states in this energy region.  The valence of Cu on
the ladder layers, $P_\ell$, is determined from the formal Cu valence
$P_{\rm av}$ as $P_\ell n_\ell + P_c n_c = P_{\rm av} (n_\ell + n_c)$
where $n_\ell$ ($n_c$) is the number of Cu atoms on the ladder (chain)
layers.  The results are summarized in Table~III.  It is noticeable
that the calculated Cu valence on the ladder layers is close to 2
while that on the chain layers shows a large deviation from 2.  It
means that the most of the holes are on the chain layers.  The Cu
valence on the chain layers slightly decreases with the Ca
substitution, indicating possible hole transfer from the chains to the
ladders.  However, the difference $0.02\sim0.05$ is smaller than the
proposed value, 0.18 for $x=0\rightarrow11$, from optical
measurement.\cite{uchida}

So far, we have ignored structure modulations and possible change of
internal atomic positions upon Ca substitution.  Since the largest
effect on charge distribution is via Madelung potential, the detailed
crystal structure is important to take into account.  We have
evaluated the effect of atomic displacements using the crystal
structures where Sr atoms are uniformly shifted in the $b$-axis
direction.  If the shift is toward the chain layers, the potential for
holes in the chain layers increases and the holes would move to the
ladder layers.\cite{mizuno} Our calculations show that the ladder Cu
valence changes about 0.06 for $\frac{b}{100}=0.13$\AA\ shift of Sr
atoms.

Another important structure modulation is displacements of O atoms in
the chain layers toward the apical positions of Cu atoms in the ladder
layers.\cite{ohta} Since O atoms in the chain layers approach positively 
charged Sr and Cu, the potential for holes on the chain layers
increases and the holes would transfer to the ladder layers.  We
have performed the calculation for Sr$_{10}$Cu$_{17}$O$_{29}$ with the
modulated structure reported in Ref.~\onlinecite{kato1}, with the
lattice constants fixed. The ladder Cu valence increases to 1.98,
compared to 1.93 for the unmodulated structure. These results indicate
the importance of structure modulation to the hole doping in the
ladders.

In summary, we have studied the electronic structures of
Sr$_{14-x}$Ca$_{x}$Cu$_{24}$O$_{41}$ within the local density
approximation. The total density of states is compared with
photoemission experiments with fairy good agreement. The energy bands
near Fermi energy are from the ladder and chain layers. They have a
pseudo one-dimensional character and the inter-ladder hoppings are
5--20\% of the intra-ladder ones. The holes are mainly on the chains
at $x=0$.  The change of the lattice constant is not enough to
transfer the holes to the ladders.  Detailed atomic displacements have
important effects on the hole distribution.

We would like to thank to O.~K.~Andersen and O.~Jepsen for providing
us the TB-LMTO-46 program. A part of numerical calculations were 
performed on the NEC SX-3/4R at the Computer Center at the Institute
for Molecular Science in Okazaki, Japan.

\onecolumn

\begin{figure}
  \epsfxsize=14.0cm \epsfbox{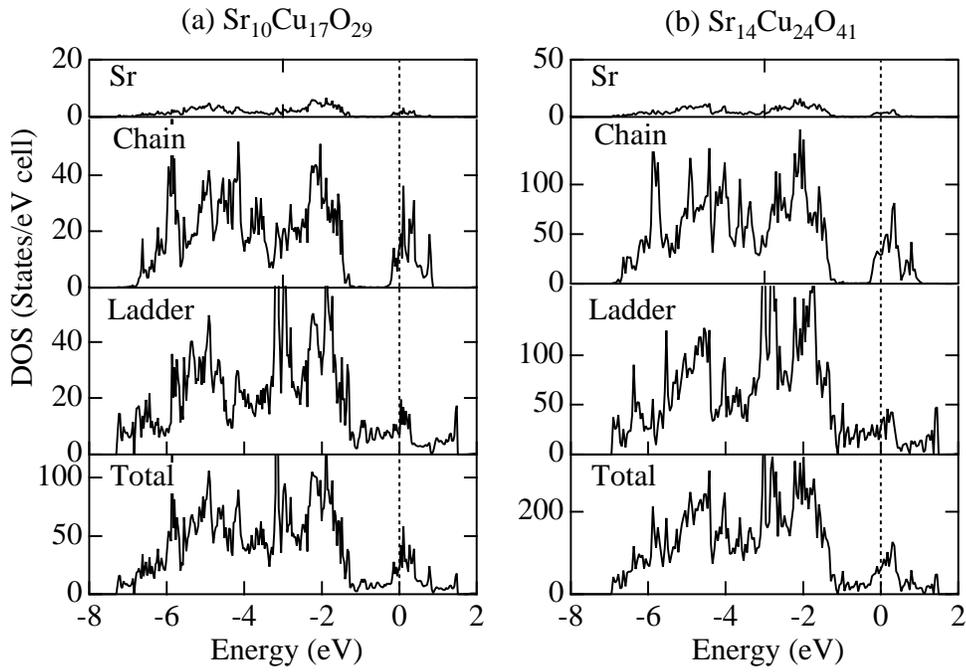}   
\caption{The total and partial density of states of
(a) Sr$_{10}$Cu$_{17}$O$_{29}$ and (b) Sr$_{14}$Cu$_{24}$O$_{41}$. }
\end{figure}

\begin{figure}
  \epsfxsize=14.0cm \epsfbox{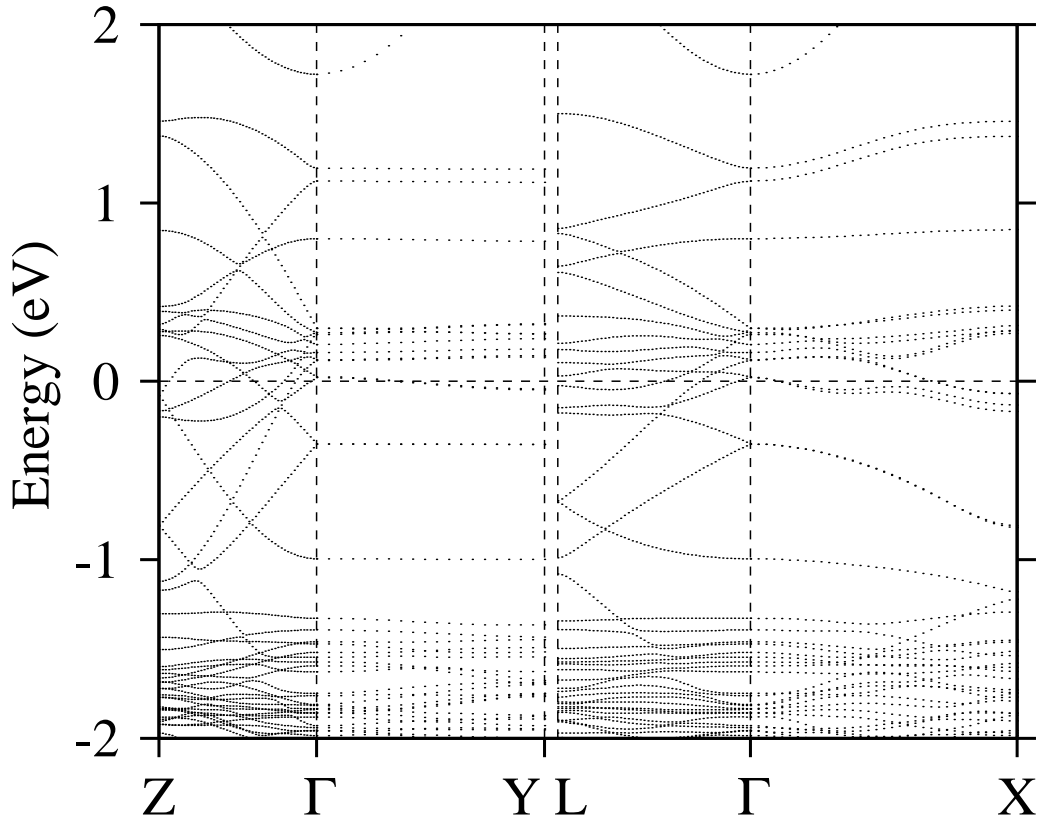}
\caption{The energy bands of Sr$_{10}$Cu$_{17}$O$_{29}$. $\Gamma$:(0,0,0),
 X:($\frac{2\pi}{a}$,0,0), Y:(0,$\frac{2\pi}{b}$,0), 
 Z:(0,0,$\frac{2\pi}{c}$), 
and L:($\frac{\pi}{a}$,$\frac{\pi}{b}$,$\frac{\pi}{c}$).}
\end{figure}

\begin{figure}
  \epsfxsize=16.0cm \epsfbox{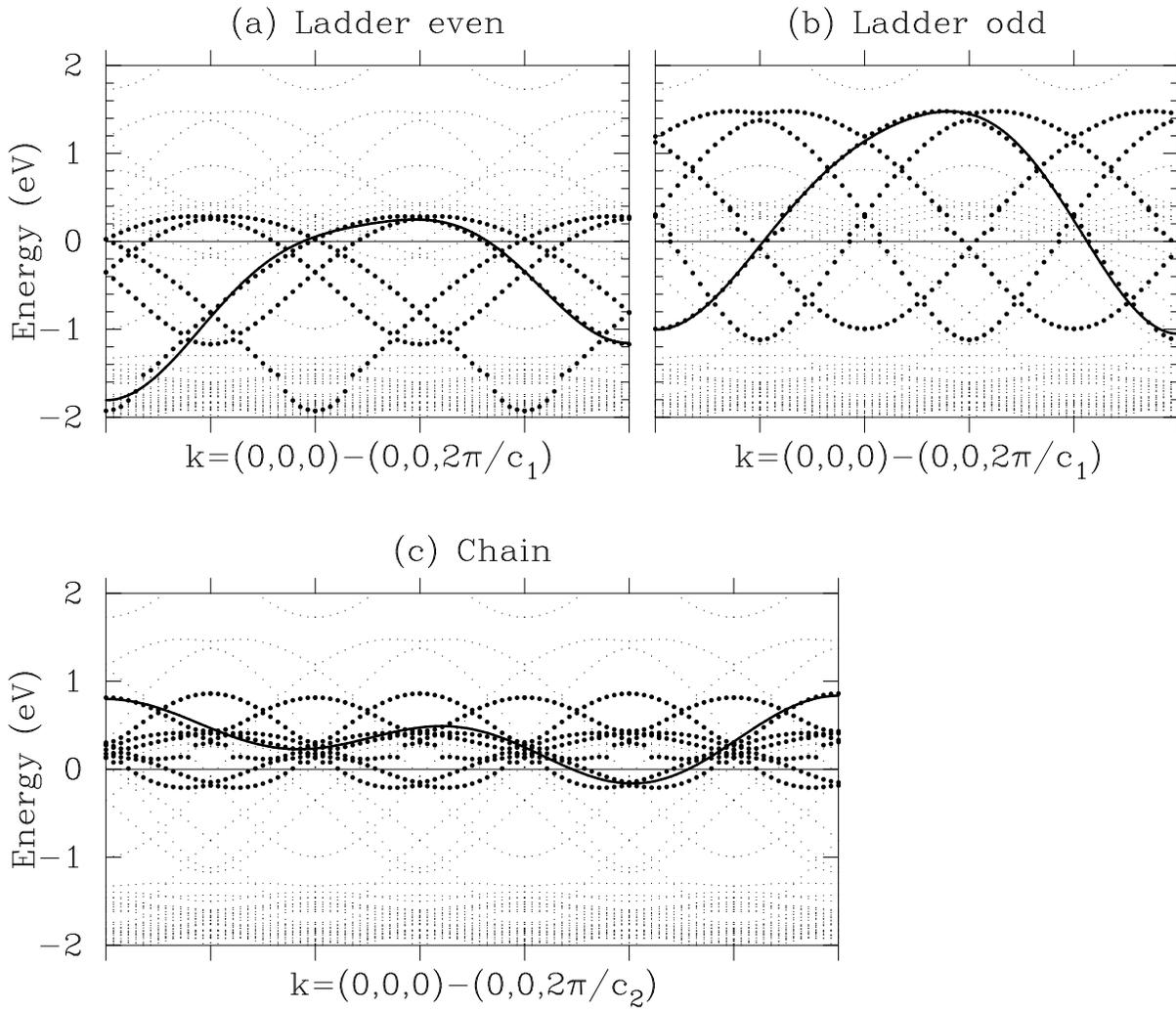}
\caption{The energy bands of Sr$_{10}$Cu$_{17}$O$_{29}$ in the
  extended Brillouin zone.  Large dots represent (a) states on ladder
  layers with even mirror symmetry, (b) those on ladder layers with
  odd mirror symmetry, and (c) those on chain layers.  Solid lines are
  numerical fit.  }
\end{figure}
\clearpage
\narrowtext
\begin{table}
  \caption{Lattice constants (\AA) used for calculations. }
  \begin{tabular}{rcccc}
    & Sr$_{10}$Cu$_{17}$O$_{24}$
    & Sr$_{14}$Cu$_{24}$O$_{41}$
    & Ca$_{10}$Cu$_{17}$O$_{24}$
    & Ca$_{14}$Cu$_{24}$O$_{41}$   \\
\tableline
$a$       &  11.46         & 11.46   & 11.46 &   11.18 \\
$b$       &  13.40         & 13.40   & 12.40 &   12.40 \\
$c$       &  19.40         & 27.65   & 19.40 &   27.10 \\
$c_1$     &  3.88          & 3.95    &  3.88 &   3.87  \\
$c_2$     &  2.77          & 2.76    &  2.77 &   2.71
\end{tabular}
\end{table}
\widetext
\begin{table}
  \caption{Tight-binding parameters (eV) of the bands near $E_F$.
    The columns with ladder(+) and ladder($-$) are for the
    ladder bands with even and odd mirror symmetry, respectively. }
  \begin{tabular}{ccccccccccccc}
    & \multicolumn{3}{c}{Sr$_{10}$Cu$_{17}$O$_{24}$} &
      \multicolumn{3}{c}{Ca$_{10}$Cu$_{17}$O$_{24}$} &
      \multicolumn{3}{c}{Sr$_{14}$Cu$_{24}$O$_{41}$} &
      \multicolumn{3}{c}{Ca$_{14}$Cu$_{24}$O$_{41}$} \\ 
    & ladder(+) & ladder($-$) &chain
    & ladder(+) & ladder($-$) &chain
    & ladder(+) & ladder($-$) &chain
    & ladder(+) & ladder($-$) &chain \\
\tableline
    $\varepsilon_0$ & $-$0.48 &\phantom{$-$}0.35& \phantom{$-$}0.34
                    & $-$0.43 &\phantom{$-$}0.43& \phantom{$-$}0.32
                    & $-$0.31 &\phantom{$-$}0.46& \phantom{$-$}0.33
                    & $-$0.33 &\phantom{$-$}0.54& \phantom{$-$}0.23\\ 
    $t_1$           &\phantom{$-$}0.42&\phantom{$-$}0.61&$-$0.09
                    &\phantom{$-$}0.44&\phantom{$-$}0.61&$-$0.12
                    &\phantom{$-$}0.41&\phantom{$-$}0.59&$-$0.09
                    &\phantom{$-$}0.44&\phantom{$-$}0.62&$-$0.11   \\ 
    $t_2$           &\phantom{$-$}0.08&\phantom{$-$}0.08&$-$0.15
                    &\phantom{$-$}0.10&\phantom{$-$}0.08&$-$0.20
                    &\phantom{$-$}0.08&\phantom{$-$}0.07&$-$0.17
                    &\phantom{$-$}0.10&\phantom{$-$}0.07&$-$0.20   \\ 
$t^{\perp}_1$       &\phantom{$-$}0.08&\phantom{$-$}0.03&$-$0.03
                    &\phantom{$-$}0.08&\phantom{$-$}0.04&$-$0.04
                    &\phantom{$-$}0.07&\phantom{$-$}0.03&$-$0.03
                    &\phantom{$-$}0.08&\phantom{$-$}0.07&$-$0.04   \\ 
$t^{\perp}_2$       &\phantom{$-$}0.00& $-$0.04 &\phantom{$-$}0.03 
                    &\phantom{$-$}0.00& $-$0.04 &\phantom{$-$}0.04 
                    &\phantom{$-$}0.00&$-$0.04  &\phantom{$-$}0.03 
                    &\phantom{$-$}0.00& $-$0.07 &\phantom{$-$}0.05 \\ 
  \end{tabular}
\end{table}
\narrowtext
\begin{table}
  \caption{Estimated Cu valence.}
  \begin{tabular}{ccccc}
    & Sr$_{10}\!$Cu$_{17}\!$O$_{24}$
    & Sr$_{14}\!$Cu$_{24}\!$O$_{41}$   
    & Ca$_{10}\!$Cu$_{17}\!$O$_{24}$
    & Ca$_{14}\!$Cu$_{24}\!$O$_{41}$   \\
    \tableline
    average     & 2.235& 2.25  & 2.235& 2.25 \\
    chain       & 2.67 & 2.64  & 2.62 & 2.62 \\
    ladder      & 1.93 & 1.97  & 1.97 & 1.99 \\
  \end{tabular}
\end{table}

\end{document}